\documentclass[a4paper,12pt]{article}

\usepackage{amsmath}
\usepackage{amsfonts}
\usepackage{graphicx}
\usepackage{color}

\numberwithin{equation}{section}
\numberwithin{figure}{section}

\newcommand{\D}{\mathrm{d}}
\newcommand\RR{\mathbb{R}}

\newcommand\EE{\mathbb{E}}

\newcommand\CC{\mathbb{C}}

\newcommand\HH{\mathcal{H}}

\newcommand\RC{\mathcal{R}}

\newcommand\btheta{\boldsymbol{\theta}}
\title{Aspects of geodesical motion with Fisher-Rao metric: classical and quantum}
\author{Florio M.\ Ciaglia$^{1,2}$, Fabio Di Cosmo$^{1,2}$, Domenico Felice$^{3,4}$,\\ Stefano Mancini$^{3,4}$, Giuseppe Marmo$^{1,2}$ and Juan M.\ P\'{e}rez-Pardo$^{2,5}$}
\date{
    \begin{flushleft}
    \small 
    $^1$ Dipartimento di Fisica, Universit\`{a} di Napoli ``Federico II'', Via Cintia Edificio 6, I--80126 Napoli, Italy.\\%
    $^2$ INFN-Sezione di Napoli, Via Cintia Edificio 6, I--80126 Napoli, Italy.\\%
    $^3$ School of Science and Technology, University of Camerino, I-62032 Camerino, Italy \\%
    $^4$ INFN-Sezione di Perugia, Via A. Pascoli, I-06123 Perugia, Italy\\
    $^5$ Departamento de Matem\'aticas, Universidad Carlos III de Madrid, Avd. de la Universidad 30, 28911 Legan\'es, Spain.\\[2ex]%
    \end{flushleft}
    \today
}

\begin{document}

\maketitle

\begin{abstract}{The purpose of this article is to exploit the geometric structure of Quantum Mechanics and of statistical manifolds to study the qualitative effect that the quantum properties have in the statistical description of a system. We show that the end points of geodesics in the classical setting coincide with the probability distributions that minimise Shannon's Entropy, i.e. with distributions of zero dispersion. In the quantum setting this happens only for particular initial conditions, which in turn correspond to classical submanifolds. This result can be interpreted as a geometric manifestation of the uncertainty principle.}
\end{abstract}

%%%%%%%%%%%%%%%%%%%%%%%%%%%%%%%%%%%%%%%%%%%%%%%%%%%%%%%%%%%%%%%%%%%%%%%%%%%%%%%%%%%%%%%%%%%%%%%%%%%%%%%%%%%%%%%%%%%%%%%%%%%%%%%%%%%%%%%%%%%%%%%%%%%%%%%%%%%%%%%%%%%%%

\section{Introduction}

In Quantum Mechanics the pure states of a system are described as elements of $\RC(\HH)$, the complex projective space of a complex separable Hilbert space $\HH$ or space of rays. From a geometric point of view, cf.\ \cite{Erc10,pepin15}, the space of rays is a K\"ahlerian manifold, i.e. it has the structure of a differentiable manifold endowed with a Riemannian metric $g$ known as the Fubini-Studi metric, a symplectic structure $\omega$ and an almost complex structure $J$ that satisfy the compatibility condition:

\begin{equation}
	g(u,v) = \omega(J u , v)\,,\quad u,v\in\mathfrak{X}(\RC(\HH))\;,
\end{equation}
where $\mathfrak{X}(\RC(\HH))$ is the space of vector fields on $\RC(\HH)$. At the level of Quantum Mechanics, the symplectic structure possesses a clear role. Indeed, the Schr\"odinger equation, as a first order evolution equation, is Hamiltonian with respect to it. In contrast, the role of the Riemannian metric is less understood. It was shown in \cite{ashtekar} that the geodesic distance between points in $\RC({\HH})$, i.e.\ between two states, with respect to the Levi-Civita connection can be associated with the transition amplitudes to go from one state to the other. Moreover, in \cite{Woo81,FN95,Brody} it is shown that there is a much deeper relationship between the Riemannian structure and the underlying statistical structure of the problem. Indeed, consider that the Hilbert space is the space of square integrable functions over a measure space $(X,\mu)$. In this case the states of the quantum system are normalised wave functions $\psi:X\to \CC$, $\int_X|\psi(x)|^2\D\mu = 1\;,$ whose moduli square represent the probability densities over the measure space $X$. Now, consider that a particular subset of $\RC(\HH)$ is parameterised by a family $M =\{\btheta = (\theta^1, \dots, \theta^m )\}\subset\RR^m$\,. Assuming that the parametrisation is one-to-one, the polar representation of the elements in $\RC(\HH)$,

\begin{equation}\label{eq:polarpsi}
	\psi(x;\btheta) = \sqrt{p(x;\btheta)}e^{i\alpha(x;\btheta)},
\end{equation}
provides an embedding of the family $M$ into $\RC(\HH)$. Here $\alpha(x;\btheta)$ is a real valued function and $p \in \mathcal{L}^1(X)$ is a probability density on the space $X$. The Hermitean tensor $h(\cdot,\cdot) =g(\cdot,\cdot) + i\omega(\cdot,\cdot)$ on the space of rays can be \emph{pulled-back} to $M$ where it takes the form
\begin{equation} \label{hermitian}
	h=\frac{1}{4} \EE_p[(d\ln p)^2]+\EE_p[(d\alpha)^2]-\EE_p[d\alpha]^2-i\ \EE_p[d\ln p\wedge d\alpha].
\end{equation}
In this expression $\EE_p[f]=\int_X dx\ p(x;\btheta) f(x)$ stands for the expectation value of the measurable function $f$. The exterior derivative and the wedge product are defined in the usual way \cite{Helgason}.
Remarkably, the hermitean tensor $h$ coincides with the classical Fisher-Rao information metric when $\D\alpha \equiv 0$, cf \cite{Marmo, Nagaoka}.
{As happens for the Fisher-Rao metric \cite{BZ06}, the probability densities have to be taken such that $p(x;\btheta)>0$ for all $\btheta \in M$ and $x\in X$. If this is not the case, the \emph{pull-back} of the metric may not be well defined.}
Hence, the parameter space $M$ associated with the polar decomposition inherits the Riemannian structure provided by Fisher-Rao metric if the quantum behaviour represented by the phase $\alpha$ vanishes. However, this will not happen in general and, in particular, the manifold may acquire the structure of a K\"ahlerian manifold. Its symmetric part is a Riemannian metric that coincides with the Fisher-Rao metric when the variation of the phase vanishes. Otherwise, there is a contribution of the phase even in the symmetric part of the hermitean tensor $h$.

It is the purpose of this article to explore the difference between the two Riemannian structures, the Fisher-Rao metric and the Fubini-Study metric. A full statistical interpretation of the latter is still to be unveiled and might help in the understanding of problems of fundamental nature in Quantum Mechanics such as the measurement process or to find more physically meaningful generalisations of the Cramer-Rao inequality, see \cite[Chapter 7]{AN00} for an introduction to this problem. {There exist generalisations of the Cramer-Rao inequality that apply in the quantum setting \cite{Helstrom76, Holevo11}. However, the interpretation of the resulting inequalities (generically) comes only after restriction to very special situations. For instance, when restricted to the classical setting where one recovers the standard Cramer-Rao bound. A study of the statistical properties of the geometric structures that are already available in the Quantum setting can lead to new and more meaningful results.}

In particular, we will study what implications does the complex phase $e^{i\alpha(x;{\btheta})}$ have in the geometric description of the problem. We will see that the classical behaviour of the system can be recovered as a totally geodesic submanifold of the quantum one. More interestingly, we find that the end points of the geodesics in the classical situations are zero dispersion states while they are never part of the geodesics nor their closures in the quantum situation. As explained in the conclusion section, this can be interpreted as a geometric manifestation of the uncertainty principle. The relation between the classical and the quantum setting that we are going to study is slightly different than the usual correspondence between classical probability distributions and diagonal mixed states. In our case the quantum situation is described only by pure states and the relation with the corresponding classical situation is given by means of  \eqref{hermitian}.\\

In order to establish these results we consider two different situations. A discrete probability space in Section \ref{sec:discprob} and a family of monovariate Gaussian distributions in Section \ref{sec:gauss}. We study the geodesic curves on the statistical manifolds so defined with respect to the given metrics. In the realm of Information Geometry there are different affine connections that play a relevant role. The most important being the so called $\alpha$-connections, cf. \cite{Barndorff83}. Since we want to compare the classical situation with the quantum one, and in the quantum setting there are no known affine connections that play such a relevant role, we will consider only the geodesics with respect to the Levi-Civita connection.
The interpretation of geodesic curves in statistical manifolds relies in the framework of inductive inference (Maximum Entropy Methods \cite{C02,C04}) and Information Geometry \cite{AN00}. An update on the knowledge of the system, through further measurements for instance, results in better estimates of the states. The geodesic joining the previous estimate with the newer one is the path that joins both points with a minimal increase of the relative entropy during the intermediate steps. Successive measurements do not need to give raise to points in the same geodesic. Therefore, repeated measurements will lead to a piecewise geodesic path on the manifold and thus the study of the endpoints of the geodesic flow on the manifold provides an understanding of what the possible optimal estimates are.

%%%%%%%%%%%%%%%%%%%%%%%%%%%%%%%%%%%%%%%%%%%%%%%%%%%%%%%%%%%%%%%%%%%%%%%%%%%%%%%%%%%%%%%%%%%%%%%%%%%%%%%%%%%%%%%%%%%%%%%%%%%%%%%%%%%%%%%%%%%%%%%%%%%%%%%%%%%%%%%%%%%%%

\section{Discrete Probability space}\label{sec:discprob}

Let us consider the statistical manifold of all two dimensional probability vectors, $\mathbb{P}^2$. In this case $X=\{0,1\}$, $p(0,\btheta) = p$ and $p(1,\btheta)= 1 - p$ with $p \in (0,1)$\,. The statistical manifold is therefore $\mathcal{M} = (0,1) \subset \mathbb{R}$\,. That is, every probability distribution is parameterised by a single real variable $p\in \left(0,1\right)$. 

Notice that since the probability space is discrete, the integrals on the continuous variable appearing in \eqref{eq:polarpsi} and \eqref{hermitian} have to be replaced by a sum.
The Fisher-Rao metric for this statistical manifold has the form
\begin{equation}
\label{FRM 1-simplex}
g=\frac{1}{p(1-p)}dp\otimes dp \,.
\end{equation}

Let us analyse the geodesic curves associated to this metric tensor. The equations of the geodesics can be written immediately:

\begin{equation}
\label{EOM 1-simplex}
    \frac{\ddot{p}}{p(1-p)} - \frac{{\dot{p}}^2(1-2p)}{2p^2(1-p)^2} = 0 \,.
\end{equation}
and the only constant of the motion is the ``kinetic energy'' term

\begin{equation}
\label{COM 1-simplex}
\frac{{\dot{p}}^2}{p(1-p)} = C \,.
\end{equation}

The diffeomorphism
\begin{equation}
\label{diffeomorphism 1-simplex}
2(p-\frac{1}{2}) = \sin y
\end{equation}
allows us to simplify the expression \eqref{COM 1-simplex}, which then assumes the form

$$
{\dot{y}}^2 = C \,.
$$

\begin{figure}[h!]
\centering
\includegraphics[height=9cm]{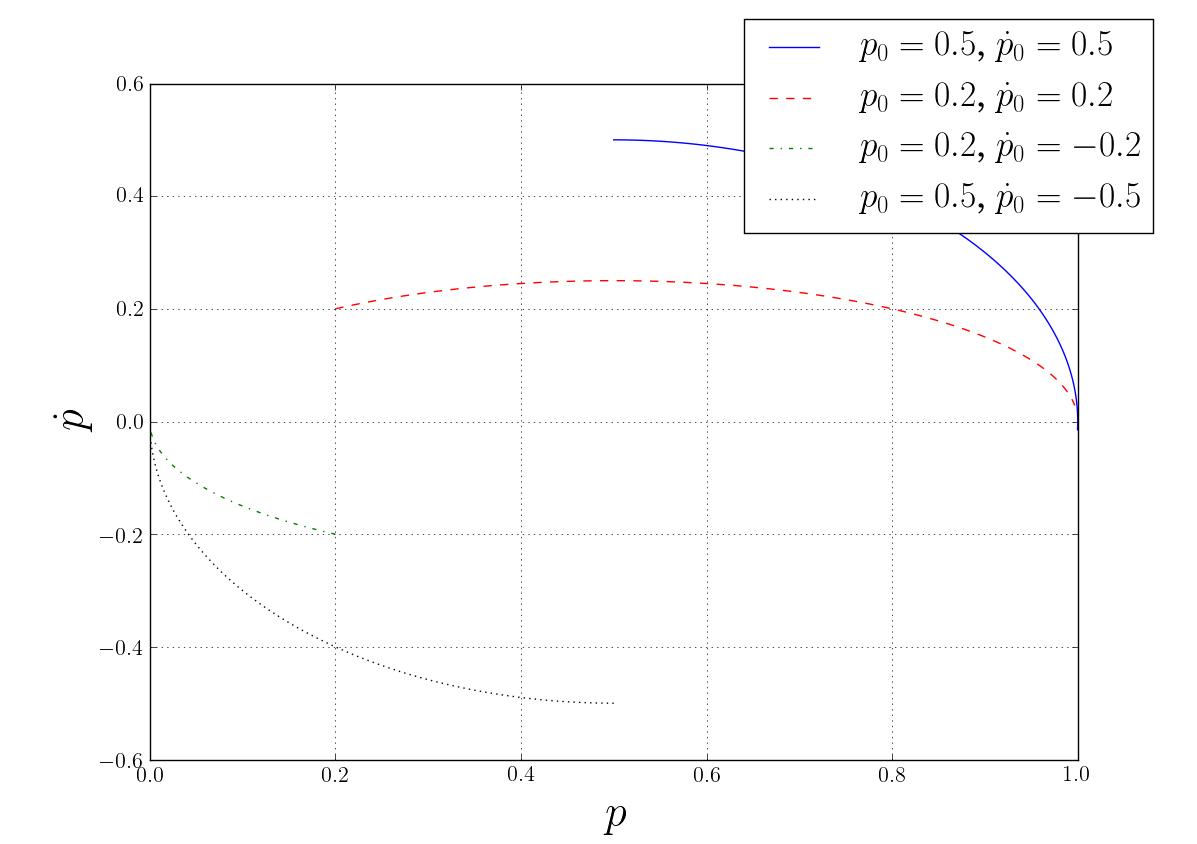}
\caption{\small{Plot of $\dot{p}$ against $p$ for different initial conditions.}}
\label{fig:orbitdiscprob}
\end{figure}

It immediately follows that, with respect to the variable $y$, solutions of the equations of motion are curves with constant velocity and all of them reach one of the two extremes of the interval. Let us incidentally note that these extreme points are reached in a finite time and therefore that the geodesic vector field is not complete. This is not a problem since the parameter of the solution has not any statistical nor dynamical interpretation. Therefore, all the geodesics associated with the Fisher-Rao metric \eqref{FRM 1-simplex} end up in the points that minimize Shannon's entropy \cite{Shannon49}. In Fig.\ \ref{fig:orbitdiscprob} one can see the numerical integration of the second order system \eqref{EOM 1-simplex}. One can see that the orbits go to the aforementioned extremal points.\\

Let us now consider the space of all pure states of a two level quantum system. This space is the Complex Projective Hilbert space $\mathbb{CP}^1$ which is diffeomorphic to the two dimensional sphere as a real manifold, cf \cite{ashtekar,Manko}. 
By referring to \eqref{eq:polarpsi} we can write a generic pure state on $\mathbb{CP}^1$ with  
$$
p(x,\btheta)=
\begin{cases}
    p,  &\; $x=0$\\
    1-p,  & \; $x=1$
\end{cases}
$$
and
$$
\alpha(x,\btheta)=
\begin{cases}
    0,  &\; $x=0$\\
    \varphi,  & \; $x=1$
\end{cases}
$$ 
with $\varphi\in[0,2\pi)$.

%Consider the following parametrisation of a pure state on $\mathbb{CP}^1$:
%$$
%\sqrt{p}|0 \rangle + \sqrt{1-p}e^{i\varphi} |1 \rangle\;,
%$$
%where $\left\lbrace |0 \rangle , |1 \rangle \right\rbrace$ forms an orthonormal basis of the two dimensional Hilbert space $\mathbb{C}^2$.
As explained in the introduction, we have to exclude the two poles of the sphere, since the \emph{pull-back} of the Fubini-Study metric is not going to be defined there. The resulting space is therefore going to be $\mathbb{CP}_0^1 \simeq \left( 0,1 \right)\times S^1 $.

With the chosen parametrisation we get

\begin{equation}
\label{FSM 2dHS}
g= \frac{1}{4p(1-p)}dp\otimes dp + p(1-p)d\varphi \otimes d\varphi \, .
\end{equation}
and the corresponding geodesic equations are
\begin{equation}\label{QEOM 2dHS}
    \frac{\ddot{p}}{4p(1-p)} - \frac{{\dot{p}}^2(1-2p)}{8p^2(1-p)^2} - \frac{1}{2}(1-2p){\dot{\varphi}}^2 = 0\;,
\end{equation}
\begin{equation}
    \frac{d}{dt}\left( p(1-p){\dot{\varphi}} \right) = 0\;.
\end{equation}

First, let us notice that the space of probability vectors $\mathbb{P}^2$ discussed previously can be recovered as any of the totally geodesic submanifolds that satisfy $\varphi = \mathrm{const}$, cf.\ \cite{Helgason}. Taking into account that $\varphi$ is a cyclic variable, there are two constants of the motion which can be used in order to get information on the solutions:
\begin{equation}\label{QCOM 2dHS}
    p(1-p)\dot{\varphi} = A 
\end{equation}
\begin{equation}
    \frac{{\dot{p}}^2}{4p(1-p)} + \frac{1}{2}p(1-p){\dot{\varphi}}^2 = C
\end{equation}

Also in this case we can perform the transformation \eqref{diffeomorphism 1-simplex} to get a simplified set of equations:

\begin{equation}\label{QCOM 2dHS II}
    \cos ^2(y)\dot{\varphi}= 4 A\;,
\end{equation}
\begin{equation}
    \frac{1}{8}{\dot{y}}^2+\frac{2A^2}{\cos ^2(y)} = C \;.
\end{equation}

We are interested in studying the motion relative to variable $p$, or equivalently relative to variable $y$, and compare the motion with the previous case. The second equation can be interpreted as the equation which defines the ``energy'' level sets of a particle with a potential $U_A(y)$, that depends on the external parameter $A$, and that has two vertical assymptotes at $y=\pm\frac{\pi}{2}$.

\begin{figure}[h!]
\centering
\includegraphics[height=9cm]{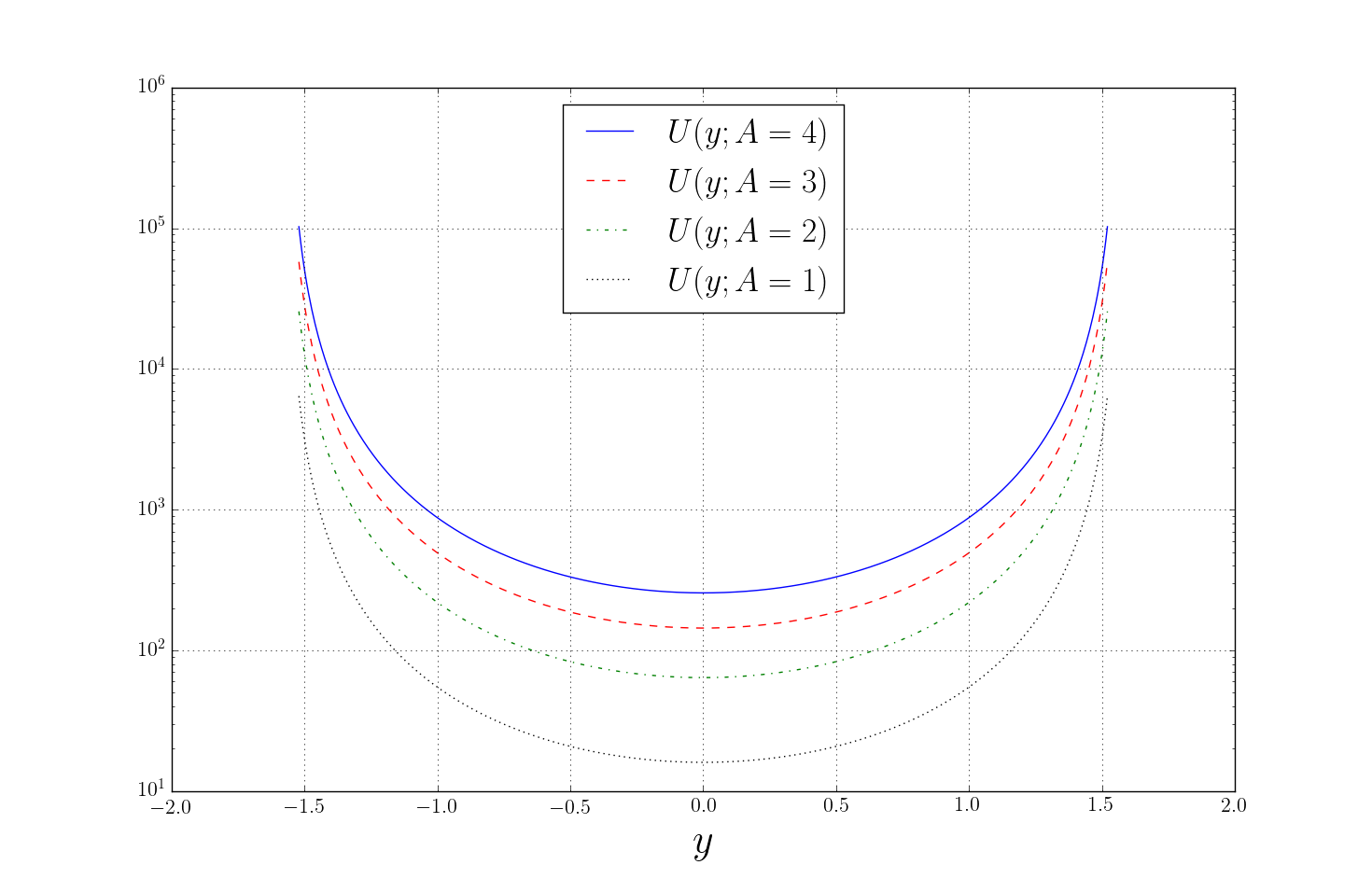}
\caption{\small{The Potential $U(y;A)$=$\frac{16A^2}{\cos ^2(y)}$ for some values of $A$. The vertical axis is in logarithmic scale.}}
\label{1-simplex}
\end{figure}

\begin{figure}[h!]
\centering
\includegraphics[height=9cm]{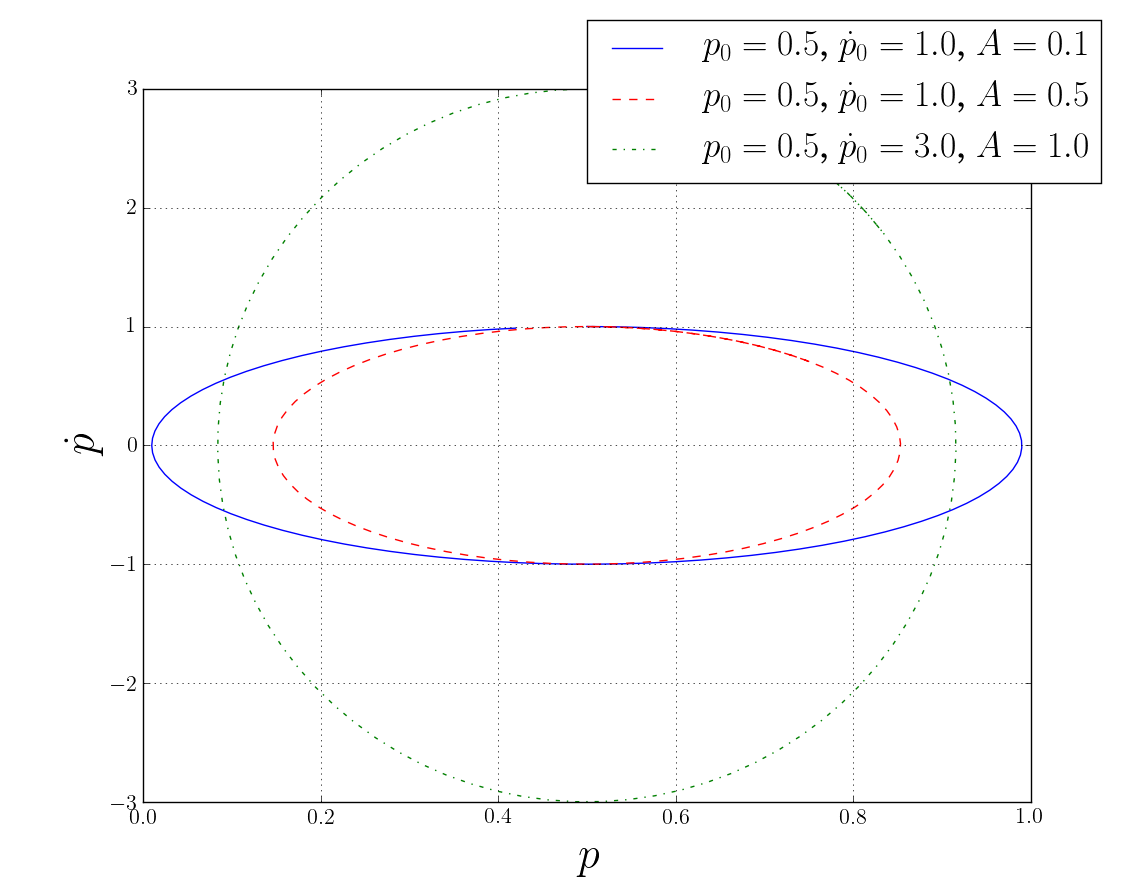}
\caption{\small{Plot of $\dot{p}$ against $p$ for different initial conditions and different values of the constant of the motion $A$. All the orbits are bounded and the extreme points $p=0$ and $p=1$ are never on them nor on their closures.}}
\label{fig:orbitdiscprobquant}
\end{figure}

In Fig.\ \ref{1-simplex} there is a plot of this potential for some values of $A$. A qualitative analysis of the solutions shows that for all $A\neq 0$ a single equilibrium point exists corresponding to the position $y_e = 0$, equivalently $p_e = \frac{1}{2}$. Moreover, for $A\neq 0$ all the admissible orbits are bounded. In particular, it must hold that
$$
-\sqrt{1-\frac{2A^2}{C}} \leq \sin(y) \leq \sqrt{1-\frac{2A^2}{C}}\;.
$$ 
In Fig.\ \ref{fig:orbitdiscprobquant} there is a plot of the numerical integration of the second order system. In this case the extreme points are not part of the closure of any orbit. In fact this holds true only if $A=0$, but this is precisely the condition for being in the totally geodesic submanifold that corresponds to the previous case. Therefore, the presence of the phase $\varphi$ affects the geodesic motion deeply, since all the orbits are bounded and the points of minimum entropy, corresponding to $y=\pm\frac{\pi}{2}$ cannot be reached. Furthermore, there is a stable fixed point, $y_e=0 $, i.e. $p=\frac{1}{2}$, which corresponds to the maximum entropy probability distribution. In addition to these geodesics there is another type of solution for the case $C=0$. These correspond to the particular case $\dot{p}_0 = 0$ and whose geodesics are just given by the points $p(t)=p_0$\,.

%%%%%%%%%%%%%%%%%%%%%%%%%%%%%%%%%%%%%%%%%%%%%%%%%%%%%%%%%%%%%%%%%%%%%%%%%%%%%%%%%%%%%%%%%%%%%%%%%%%%%%%%%%%%%%%%%%%%%%%%%%%%%%%%%%%%%%%%%%%%%%%%%%%%%%%%%%%%%%%%%%%%%

\section{Gaussian Probability Space}\label{sec:gauss}

We will perform now a similar analysis on a different statistical model. We will consider the set of Gaussian probability distributions over the real line. Every distribution can be parametrised by points of a two dimensional manifold $\mathcal{N} = \mathbb{R}\times \mathbb{R}_+$, namely:
\begin{equation}\label{eq:gauss}
    \mathcal{N} \ni (\mu, \sigma) \; \mapsto \; p(x; \mu, \sigma) = \frac{e^{-\frac{(x-\mu)^2}{2\sigma^2}}}{\sigma \sqrt{2\pi}} \;.
\end{equation}
The Fisher-Rao metric for this manifold of Gaussian probability distributions becomes
\begin{equation}\label{eq:fisherrao}
    g_{\mathcal{N}} = \frac{1}{4\sigma^2} \left( d\mu \otimes d\mu + 2 d\sigma \otimes d\sigma \right) \,,
\end{equation}
and the associated geodesic equations are
\begin{equation}\label{EOM gaussian}
    \frac{d}{dt}\left( \frac{\dot{\mu}}{2\sigma ^2} \right) = 0\;,
\end{equation}
\begin{equation}
    \frac{d}{dt}\left( \frac{\dot{\sigma}}{\sigma ^2} \right) +\frac{({\dot{\mu}}^2+2{\dot{\sigma}}^2)}{2\sigma ^3} = 0\;.
\end{equation}
There are two constants of the motion. The first one is the Lagrangian itself and the second one is the momentum associated to the cyclic variable $\mu$. Again, in order to get simplified expressions, let us consider the following diffeomorphism
\begin{equation}\label{diffeo gaussian}
    \sigma = e^y\;,
\end{equation}
and $\mu \to \mu$. Having introduced these new coordinates, the constants of the motion can be written as
\begin{equation}\label{COM gaussian}
    \frac{\dot{\mu}}{2e^{2y}} = A\;,
\end{equation}
\begin{equation}\label{eq:energygaussian}
    \frac{1}{2}\dot{y}^2 + A^2 e^{2y} = C\;.
\end{equation}
By means of a qualitative analysis of the second equation, we can notice that the potential $ U(y) = 2A^2e^{2y}$ does not possess any minimum. Equation \eqref{eq:energygaussian} then shows that every orbit is bounded from above. In particular, whatever the initial conditions are, the motion will reach a maximum value 
$$y_m = \log\left( \frac{\sqrt{C}}{|A|} \right)\;,$$
equivalent to
$$\sigma_m = \frac{\sqrt{C}}{|A|}\;,$$
and then will bounce back. There is a limiting point for all these solutions, $y \rightarrow -\infty$, which corresponds to $\sigma \to 0$. Also $\dot{\mu}$ must tend to zero in the same limit and therefore all the geodesic curves tend to the ``delta'' probability distribution with zero variance. Its support, which corresponds to the final value of $\mu$, is fixed according to the initial conditions. This result is analogous to the one recovered for the discrete probability space at Section \ref{sec:discprob}. Indeed, the limiting points for the geodesics are those which correspond to the probability distributions that minimize Shannon's entropy \cite{Shannon49}.\\ 

Let us now consider a quantum counterpart of the statistical manifold just introduced. Let $\mathcal{R}(\mathcal{H})$ be the infinite dimensional Projective Hilbert space associated to the Hilbert space $\mathcal{H} = \mathcal{L}^2(\mathbb{R},dx)$, where $dx$ is the Lebesgue measure on the real line. We can select a family of pure states according to the following injective map:
\begin{equation}
    \mathcal{M}=\mathbb{R}^2\times \mathbb{R}_+ \ni \theta = (\mu,\alpha,\sigma) \mapsto \psi(x;\theta) = \sqrt{\frac{e^{-\frac{(x-\mu)^2}{2\sigma^2}}}{\sigma \sqrt{2\pi}}}e^{i\alpha x} \in \mathcal{H} \,.
\end{equation}
Comparing with \eqref{eq:polarpsi} this means that $$p(x,\btheta) = {\frac{e^{-\frac{(x-\mu)^2}{2\sigma^2}}}{\sigma \sqrt{2\pi}}}\,,\quad \alpha(x,\btheta) = \alpha x\;.$$
In other words, to any point of the manifold $\mathcal{M}$ it corresponds a probability amplitude whose associated probability distribution in the position representation is a Gaussian. As explained in the introduction, the \emph{pull-back} of the hermitean tensor on $\mathcal{R}(\mathcal{H})$ defines two tensors on the manifold $\mathcal{M}$, see \eqref{hermitian}, one being symmetric and the other anti-symmetric. In this case these tensors have the form
\begin{equation}\label{FSM gaussian}
    g_\mathcal{M} = \frac{1}{4\sigma^2} \left( d\mu \otimes d\mu + 2 d\sigma \otimes d\sigma \right) + \sigma^2 d\alpha \otimes d\alpha\;;\quad \omega = d\alpha \wedge d\mu\;.
\end{equation}
The symmetric part, which defines a Riemannian tensor, is a quantum ``extension'' of the classical Fisher-Rao metric, cf. \cite{Marmo}. One can see the similarity with the Fisher-Rao metric, see \eqref{eq:fisherrao}, on the statistical manifold defined by the monovariate Gaussian model. 
In order to make a comparison with the previous analysis, let us consider the geodesic equations of the metric $g_\mathcal{M}$ of \eqref{FSM gaussian}:
\begin{equation}\label{QEOM gaussian}
    \frac{d}{dt}\left( \frac{\dot{\mu}}{2\sigma ^2} \right) = 0\;,\\
\end{equation}
\begin{equation}
    \frac{d}{dt}\left( 2\sigma ^2 \dot{\alpha} \right)= 0\;,\\
\end{equation}
\begin{equation}
    \frac{d}{dt}\left( \frac{\dot{\sigma}}{\sigma ^2} \right) +\frac{({\dot{\mu}}^2+2{\dot{\sigma}}^2)}{2\sigma ^3} - 2\sigma {\dot{\alpha}}^2 = 0 \,.
\end{equation}
As in the case for the discrete probability space of the previous section, we have that the manifold $\mathcal{N}$, associated to the space of the classical probability densities, is a totally geodesic submanifold of $\mathcal{M}$ characterised by $\alpha = \mathrm{const}$. 

In order to analyse the behaviour of the geodesics let us perform again the diffeomorphism \eqref{diffeo gaussian}. Then we can write the equations defining the constants of the motion:

\begin{equation}\label{QCOM gaussian}
    \frac{\dot{\mu}}{2e^{2y}} = A\;,
\end{equation}
\begin{equation}
    2e^{2y} \dot{\alpha}= B\;,
\end{equation}
\begin{equation}\label{eq:Energyorbits}
    \frac{1}{2}\dot{y}^2 + A^2 e^{2y} + \frac{B^2}{2e^{2y}} = C\;.
\end{equation}

\begin{figure}[h!]
\centering
\includegraphics[height= 9cm]{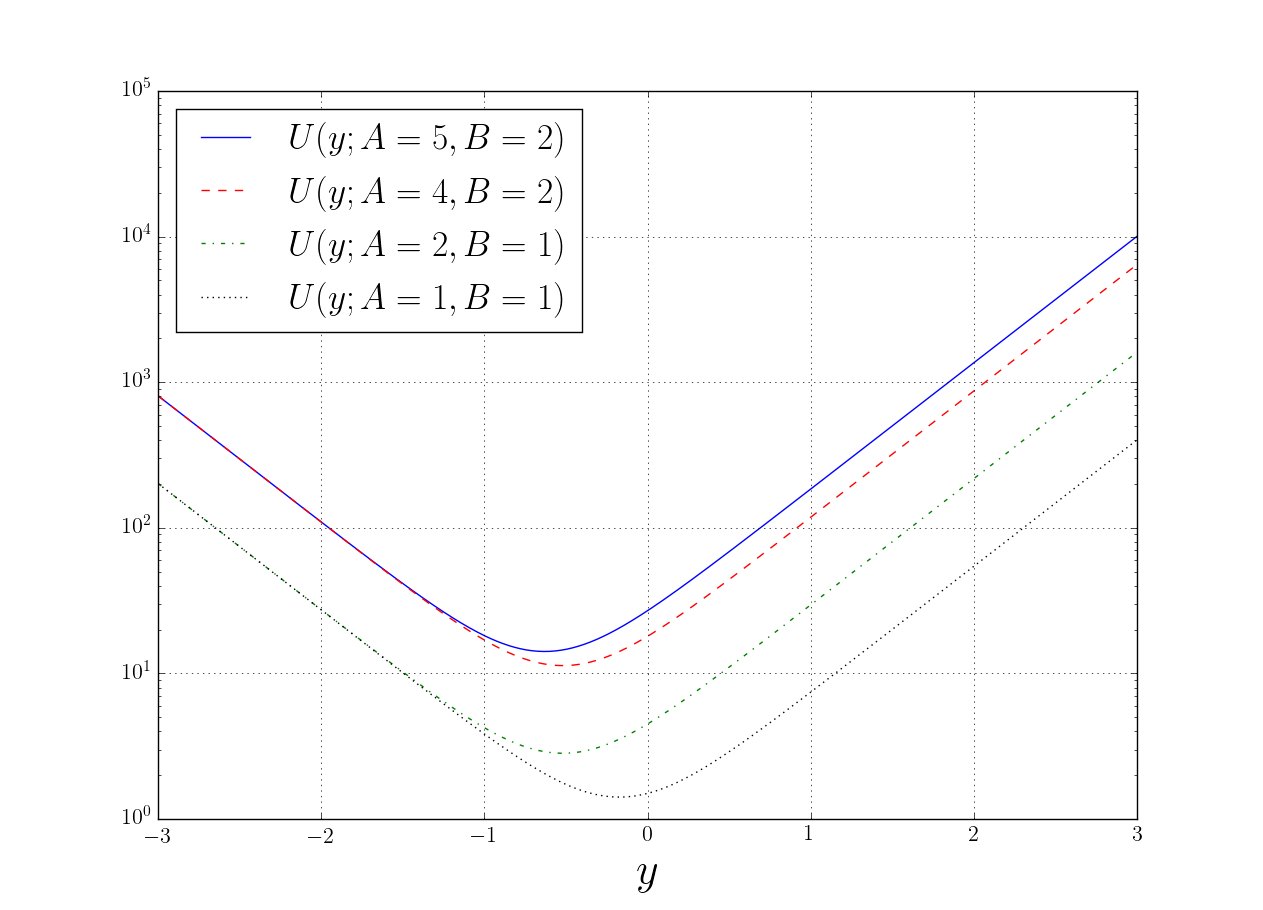}
\caption{\small{The Potential $U(y;A,B)=A^2 e^{2y} + \frac{B^2}{2e^{2y}}$ for some values of $A$ and fixed $B=1$. The vertical axis is in logarithmic scale. The behavior of the potential for $y \rightarrow -\infty$ is determined only by the value of $B$.}}
\label{gaussian}
\end{figure}

\begin{figure}[h!]
\centering
\includegraphics[height=8.5cm]{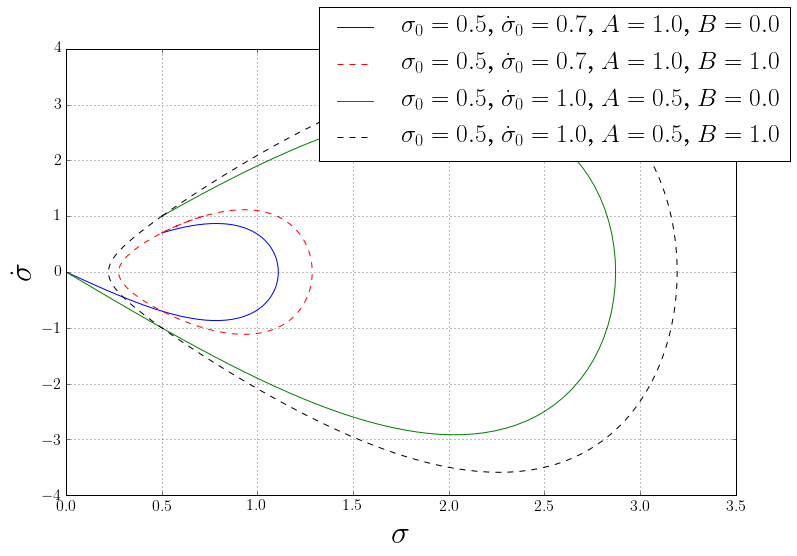}
\caption{\small{Plot of $\dot{\sigma}$ against $\sigma$ for different initial conditions and different values of the constants of the motion. The dashed lines represent the situation with $B\neq0$. The continuous lines represent the curves in the totally geodesic submanifold. The value of $C$ is determined in each case from \eqref{eq:Energyorbits} by the values of A and B and the initial data.}}
\label{fig:gaussquant}
\end{figure}

By looking at the third equation, one can notice that the parametric potential 
$$
U(y; A,B) = A^2 e^{2y} + \frac{B^2}{2e^{2y}}\;,
$$
possesses a minimum for $y_e = \frac{1}{4}\log \left( \frac{B^2}{2A^2} \right)$, as can be seen from the plots in Fig \ref{gaussian}. Therefore, all the orbits for $B\neq 0$ are bounded. For the variable $\sigma$ the bounds read:
$$
 \frac{C - \sqrt{C^2-2A^2B^2}}{2A^2}  \leq \sigma^2 \leq \frac{C + \sqrt{C^2-2A^2B^2}}{2A^2} \,.
$$ 
First notice that the situation for the statistical manifold built from the monovariate Gaussian model is recovered for the particular case $B=0$. This in turn corresponds to the case $\alpha = \mathrm{const}.$ So again, the classical geometric situation corresponding to the Fisher-Rao metric is recovered as a totally geodesic submanifold. In the first case the limiting points of the geodesics are the Dirac-delta distibutions, i.e. probability densities that minimise Shannon's entropy in the given model. In this latter situation, those limiting points are only available if one starts in the totally geodesic submanifold, otherwise they are forbidden. In Fig.\ \ref{fig:gaussquant} there is a plot comparing these two situations for different initial data and values of the constants of the motion. As can be seen, for $B\neq0$ the orbits bounce between the respective extreme values for $\sigma$ while for $B=0$ they tend to the limiting point $\sigma=0$.

Again, as in the previous section, the geodesics of the quantum Fisher-Rao metric have a completely different behaviour with respect to the corresponding quantum motion. The main difference consisting in the fact that the probability density with minimum Shannon's entropy is forbidden\footnote{Since we are talking about probability densities it makes sense to compute Shannon's entropy even if we are in a quantum setting.}. It remains to say that there is a third special kind of orbit which correspond to the situation $A=0$. These geodesics exist in both, the manifold $\mathcal{M}$ and in its totally geodesic submanifold $\mathcal{N}$. As can be checked, in either case, these geodesics have as limiting point the distribution corresponding to the limit $\sigma \to \infty$\;, while the mean value of the Gaussian is $\mu=const$\,.  This can be interpreted as the analog of the fixed points obtained in Section \ref{sec:discprob} for the case $C=0$. 

%%%%%%%%%%%%%%%%%%%%%%%%%%%%%%%%%%%%%%%%%%%%%%%%%%%%%%%%%%%%%%%%%%%%%%%%%%%%%%%%%%%%%%%%%%%%%%%%%%%%%%%%%%%%%%%%%%%%%%%%%%%%%%%%%%%%%%%%%%%%%%%%%%%%%%%%%%%%%%%%%%%%%

\section{Conclusions}

We have performed a comparative analysis of the geodesics in two situations. The first situation corresponds to the statistical manifold associated to a discrete probability space and its quantum counterpart, a finite level quantum system. The second situation corresponds to the statistical manifold associated to a monovariate Gaussian model while its quantum counterpart is given by an embedding of Gaussian wave packets  into the space of rays $\mathcal{R}(\mathcal{H})$. The geodesics compared are respectively those arising from the Fisher-Rao metric with those arising from the \emph{pull-back} of the Fubini-Study metric. We have observed that the classical situations are naturally described in this context as totally geodesic submanifolds of the respective quantum counterparts. Interestingly, in the classical scenarios the limiting points for the geodesics turn out to be those points that represent probability distributions which minimise Shannon's Entropy, i.e. $p=1$ or $p=0$ in the discrete probability space and $\sigma=0$ in the Gaussian model. This situation is forbidden in the quantum counterparts, except for those initial conditions that coincide with the totally geodesic submanifolds. This amounts to say that the quantum contribution to the problem manifests itself by preventing to achieve states of zero dispersion. Due to the nature of the examples chosen, which are simple but general enough, we expect that this will be the general case. An argument in favour of this consideration is that this can be seen as a manifestation of the uncertainty relations. Indeed, provided a pair of observables that do not commute one has that
$$\Delta x \Delta p \geq \mathbb{E} [xp - px] \,.$$
Hence, one can take $\Delta x$ as small as needed at the price of enlarging $\Delta p$. However, the value $\Delta x = 0$ is forbidden. This is completely analogous to the situation that we encountered, where $\Delta x = \sigma$ and $\Delta p = \frac{1}{2\sigma}$. Therefore this behaviour can be interpreted as a geometric manifestation of the uncertainty principle. While the choice of $\alpha(x,\btheta)$ was particular in the Gaussian model, it was general in the case of the two-level system. This supports the generality of the derived results.

The geometric approach presented here could be applied to investigate further and shed some light on the relation between classical and quantum correlations. For instance one could consider multipartite systems and the appearance of entanglement. We leave such investigations for future work.

%%%%%%%%%%%%%%%%%%%%%%%%%%%%%%%%%%%%%%%%%%%%%%%%%%%%%%%%%%%%%%%%%%%%%%%%%%%%%%%%%%%%%%%%%%%%%%%%%%%%%%%%%%%%%%%%%%%%%%%%%%%%%%%%%%%%%%%%%%%%%%%%%%%%%%%%%%%%%%%%%%%%%

\section*{Acknowledgements}
{G.\ Marmo would like to acknowledge the support provided by the Banco de Santander-UC3M ``Chairs of Excellence'' Programme 2016-2017. J.M.\ P\'{e}rez-Pardo is partly supported by the Spanish MINECO grant MTM2014-54692-P and QUITEMAD+, S2013/ICE-2801.}

%%%%%%%%%%%%%%%%%%%%%%%%%%%%%%%%%%%%%%%%%%%%%%%%%%%%%%%%%%%%%%%%%%%%%%%%%%%%%%%%%%%%%%%%%%%%%%%%%%%%%%%%%%%%%%%%%%%%%%%%%%%%%%%%%%%%%%%%%%%%%%%%%%%%%%%%%%%%%%%%%%%%%


\begin{thebibliography}{99}


\bibitem{AN00}
S. Amari, and H. Nagaoka. Methods of Information Geometry,
Oxford University Press (2000).

\bibitem{ashtekar}
{Ashtekar, A. and Schilling, T.A.}.
{Geometrical Formulation of Quantum Mechanics.}
{\it Erwin Schr\"{o}dinger International Institute for Mathematical Physics, Vienna} {(1997)}

\bibitem{Barndorff83}
O.E. Barndorff-Nielsen and P. Blaesild, \textit{Exponential  models  with
affine  dual  foliations}, Annals of Statistics {\bf 11}, 753-769 (1983).

\bibitem{BZ06}
I. Bengtsson and K. Zyczkowski. \textit{Geometry of Quantum States}. Cambridge University Press, Cambridge 2006.

\bibitem{Brody}
D.C. Brody, L.P. Hughston,   \textit{Geometrisation of statistical mechanics}, Proceedings of the Royal
Society London A {\bf 455}, 1683--1715  (1999).

\bibitem{pepin15}
J.F.\ Carinena, A.\ Ibort, G.\ Marmo and G.\ Morandi.
\textit{Geometry from Dynamics, Classical and Quantum.}
Springer Netherlands (2015).

\bibitem{C02}
A. Caticha. \textit{Entropic dynamics.} R.L. Fry(Ed.), Bayesian Inference and Maximum Entropy Methods in Science and Engineering, in: AIP COnf. Proc., {\bf 617},  p.302 (302).

\bibitem{C04} 
A. Caticha and R. Preuss, \textit{Maximum entropy and data analysis: Entropic prior distributions}, Physical Review E \textbf{70}, 046127 (2004).

\bibitem{Erc10}
E. Ercolessi, G. Marmo, and G. Morandi. \textit{From the equations of motion to the canonical commutation relations}, La
rivista del Nuovo Cimento della Societa` Italiana di Fisica, {\bf 33} 8--9 (2010).

\bibitem{Marmo}
P. Facchi, R. Kulkarni, V.I. Man'ko, G. Marmo, E.C.G. Sudarshan, F. Ventriglia, \textit{Classical and quantum Fisher information in the geometrical formulation of quantum mechanics}, Physics Letters A {\bf 374}, 0375-9601 (2010).

\bibitem{FN95}
{A.\ Fujiwara and H.\ Nagaoka}
\textit{Quantum Fisher metric and estimation for pure state models.}
{Phys. Lett. A {\bf 201}, 119--124 (1995)}

\bibitem{Helgason}
S. Helgason. Differential Geometry, Lie Groups, and Symmetric Spaces, Academic Press (1078).

\bibitem{Helstrom76}
C.W.\ Helstrom.
\textit{Quantum Detection and Estimation Theory.} Academic Press New York (1976).

\bibitem{Holevo11}
A.\ Holevo.
\textit{Probabilistic and Statistical Aspects of Quantum Theory.} Publications of the Scuola Normale Superiore. Springer Basel (2011).

\bibitem{Manko}
V.I. Man'ko, G. Marmo, F. Zaccaria, E.C.G. Sudarshan, \textit{Differential geometry of density states}, Reports on Mathematical Physics {\bf 55}, 405--422 (2005).

\bibitem{Nagaoka}
H. Nagaoka, \textit{On Fisher information on quantum statistical models},
Asymptotic Theory of Quantum Statistical Inference, 113--124 (2005).

\bibitem{Shannon49}
Shannon, C.E., Weaver, W.  \textit{The Mathematical Theory of Communication}, University of Illinois Press. (1949) 

\bibitem{Woo81}
{W.K. Wootters}
\textit{Statistical distance and Hilbert space}
{Phys. Rev. D {\bf 23}(2), 357--362 (1981)}


\end{thebibliography}
\end{document}